\documentclass[english,3p,twocolumn,sort&compress]{elsarticle}
\usepackage[T1]{fontenc}
\usepackage{babel}

\usepackage{units}
\usepackage{textcomp}
\usepackage{amsmath}
\usepackage{graphicx}
\usepackage{amssymb}
\usepackage[unicode=true, 
 bookmarks=true,bookmarksnumbered=false,bookmarksopen=false,
 breaklinks=false,pdfborder={0 0 1},backref=false,colorlinks=false]
 {hyperref}

\makeatletter

\newcommand{\lyxmathsym}[1]{\ifmmode\begingroup\def\b@ld{bold}
  \text{\ifx\math@version\b@ld\bfseries\fi#1}\endgroup\else#1\fi}

\journal{Physics Letters A}

\makeatother

\begin{document}

\title{A lower bound for the velocity of quantum communications in the preferred
frame.}

\author[Liceo]{B. Cocciaro\corref{cor1}}

\ead{b.cocciaro@comeg.it}

\author[Dipartimento1]{S. Faetti}

\ead{faetti@df.unipi.it}

\author[Dipartimento2]{L. Fronzoni}

\ead{fronzoni@df.unipi.it}

\cortext[cor1]{Corresponding author}

\address[Liceo]{Liceo Scientifico XXV Aprile, via Milano 2, 56025 Pontedera (Pi)
Italy}

\address[Dipartimento1]{Dipartimento di Fisica and Polylab of INFM, Largo B. Pontecorvo,
56123 Pisa, Italy}

\address[Dipartimento2]{Dipartimento di Fisica and INFMCRS-Soft, Largo B. Pontecorvo, 56123
Pisa, Italy}
\begin{abstract}
An EPR experiment with polarized entangled photons is performed to
test the Eberhard model. According to the Eberhard model, quantum
correlations between space-like separated events are due to a superluminal
communication signal propagating in a preferred frame. The coincidences
between entangled photons passing through two polarizers aligned along
a East-West axis are measured as a function of time during 21 sidereal
days. No deviation from the predictions of the Quantum Theory is observed.
Tacking into account for the experimental uncertainties, we infer
that, if a preferred frame for superluminal signals exists which moves
at velocity \textbf{\emph{$\vec{v}$ }}with respect to the Earth,
the modulus of the velocity of quantum communications in this frame
has to be greater than $v_{t}\simeq0.6\cdot10^{4}\, c$ for $v<0.1\, c$
and for any arbitrary direction of \textbf{\emph{$\vec{v}$}}.\end{abstract}
\begin{keyword}
Optical EPR experiments \sep Preferred frame \sep Tachyons 
\end{keyword}
\maketitle

\section{Introduction}

The non-local character of Quantum Mechanics (\textit{QM}) has been
object of a great debate starting from the famous Einstein-Podolsky-Rosen
(\textit{EPR}) paper \citep{EPR}. Consider, for instance, a quantum
system made by two photons 1 and 2 that are in the polarization entangled
state 

\begin{equation}
|A>=\frac{1}{\sqrt{2}}\left(|H,H>+e^{i\phi}|V,V>\right)\label{eq:1}\end{equation}
where \textit{H} and \textit{V} stand for horizontal and vertical
polarization, respectively, and $\phi$ is a constant phase coefficient.
The two photons propagate in space far away one from the other and
the polarization of the two photons is measured at a given time. According
to \textit{QM}, a measurement of horizontal polarization of one of
the entangled photons leads to the collapse of the entangled state
to $|H,H>$, then, also the other photon must collapse to the horizontal
polarization whatever is its distance from the previous photon. This
behavior\textit{ }suggests the existence of a sort of {}``action
at distance'' qualitatively similar to that between electric charges
and masses that was introduced before the advent of the local Maxwell
electromagnetic theory and the local Einstein General Relativity theory.
Many physicists are unsatisfied of the non-local character of \textit{QM}
and alternative local models based on hidden variables have been suggested.
As shown by Bell \citep{Bell} and other authors \citep{key-3,key-4},
local hidden variables theories must satisfy some inequalities that
are not satisfied by \textit{QM.} Many experiments have been performed
for checking these inequalities \citep{key-5,key-6,key-7,key-8,key-9,key-10,key-11,key-12,key-13}
and the validity of \textit{QM} versus local hidden variables theories
has been always found. Although the locality loophole and the detection
loophole have not yet completely closed using a single experimental
apparatus \citep{Genovese}, experiments have continuously converged
to closing both the locality loophole \citep{key-6,key-7,key-10,key-11}
and the detection loophole \citep{key-8,key-12}. Then, it seems to
us to be reasonable to think that hidden variables alone cannot fully
justify \emph{QM} correlations of entangled particles. In this case,
alternative local explanations of \emph{QM} correlations could be
possible assuming some communication between entangled particles \citep{Eberhard,Bohm}.

Indeed, as shown by Eberhard \citep{Eberhard}, a realistic local
model of \textit{QM} can be obtained assuming that some superluminal
communication (quantum tachyon) propagating in a preferred frame (\emph{PF})
can occur in a quantum system. Tachyons are known to lead to causal
paradoxes \citep{Moller} (the present in a given point can be affected
by the future in the same point), but it can be shown that no causal
paradox arises if they propagate in a preferred frame where the tachyon
velocity $v_{t}=\beta_{t}c$ ($\beta_{t}>1$) is the same in all directions
(see, for instance, section 3.1 of \citep{Liberati}). According the
Eberhard model, if two photons are in the entangled state of eq.\eqref{eq:1},
when polarization of one photon is measured and it collapses in the
horizontally oriented state, then a tachyon is sent to the other entangled
photon that collapses to the horizontally oriented state only after
this communication has been received. Therefore, the standard quantum
correlations for the polarization measurements of entangled photons
can be recovered only if it has been communication between the two
photons. If there has been not sufficient time for this communication,
a lack of quantum correlation occurs. Of course, the predictions of
the Eberhard model coincide entirely with those of \textit{QM} if
$v_{t}\rightarrow\infty$, then no experiment satisfying \textit{QM}
can invalidate this model but it can only fix a lower bound for the
tachyon velocity $v_{t}$.

A long-distance (10.6 km) \textit{EPR} experiment has been performed
by Scarani et al. \citep{Scarani} using energy-time entangled photons
to establish a lower bound for velocity $v_{t}$ of quantum communications.
The experimental results were analyzed under the assumption that the
\textit{PF} is the frame of cosmic microwave background radiation.
With this assumption, they obtained a lower bound $v_{min}=1.5\cdot10^{4}\, c$.
More recently similar measurements have been performed by Salart et
al. \citep{Salart} improving some features of the previous experiment
and using detectors aligned close to East-West direction (at angle
$\alpha=5.8\lyxmathsym{\textdegree}$). In such a way the authors
were able to find a higher value of the lower bound of the tachyon
velocity for any direction of velocity \textbf{\emph{$\vec{v}$}}
of the \emph{PF}. In this Letter we report the results of a small
distance ($2\, m$) \textit{EPR} experiment with polarization entangled
photons. The polarization measurements on the two entangled photons
are made in two points \textit{A} and \textit{B} aligned along the
East-West direction. Whatever is the orientation of velocity \textbf{\emph{$\vec{v}$}}
of the preferred frame, an appreciable disagreement with quantum predictions
would be expected if the tachyon velocity in the \textit{PF} is lower
than a minimum value $\beta_{t,min}c$ (see eq.\eqref{eq:11-2}).
Since we do not find any deviation from the predictions of \textit{QM},
we infer that possible tachyons velocity exceed $\beta_{t,min}c$
in agreement with the results in refs. \citep{Scarani} and \citep{Salart}.
In Section \ref{sec:Conditions-for-the} we discuss what are the conditions
that make possible the experimental observation of a lack of quantum
correlation. In Section \ref{sec:--Experimental-apparatus} we discuss
the main features of our apparatus and the main experimental uncertainties.
In Section \ref{sec:--Experimental-results} we report our results
and the conclusions.

\section{\label{sec:Conditions-for-the}Conditions for the lack of quantum
correlation of entangled photons.}

\begin{figure}[tbh]
\centering{}\includegraphics[width=5.5cm,height=5cm]{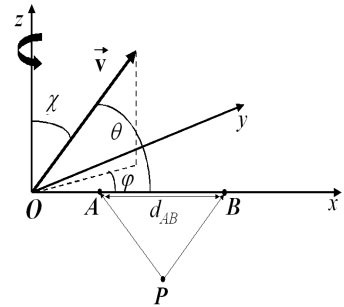}\caption{\textit{\label{fig:ExpMeth}P}: position of the source of the entangled
photons; \textit{A,B}: positions of the polarizers; \textbf{\emph{$\vec{v}$}}:
velocity of the tachyons preferred frame (\textit{PF}); \textit{x}:
East-West axis in the laboratory; \textit{z}: rotation axis of the
Earth. }

\end{figure}

The main features of the experimental method are schematically drown
in fig.\ref{fig:ExpMeth}. Two entangled photons are generated at
a point \emph{P} and meet two polarizers at points \emph{A} and \emph{B}
at distance $d_{AB}=x_{B}-x_{A}>0$ along a \emph{x}-axis of the laboratory
frame oriented along the East-West direction.

\begin{figure}[tbh]
\centering{}\includegraphics[scale=0.3]{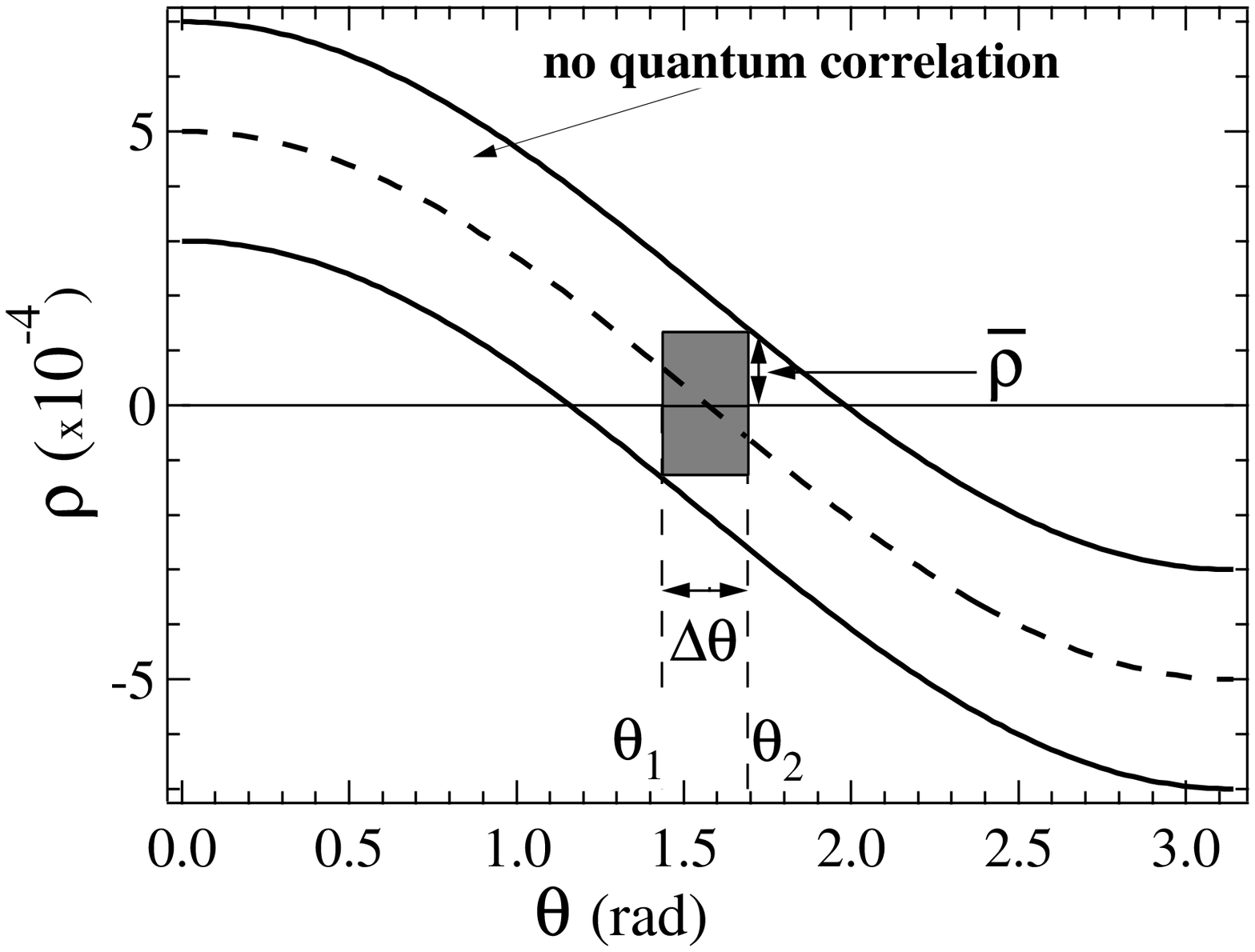}\caption{\label{fig:DeltaSuTheta}The full curves represent functions $-\nicefrac{c}{v_{t}^{-}\left(\theta\right)}$
(lower curve) and $\nicefrac{c}{v_{t}^{+}\left(\theta\right)}$ (upper
curve), respectively, for $\beta=0.5\cdot10^{-3}$ and $\beta_{t}=0.5\cdot10^{4}$.
The broken curve represents function $y=\beta\cos\theta$. The region
between the full curves corresponds to the region where no quantum
correlation occurs (eq.\eqref{eq:6}).}

\end{figure}

The \textit{z}-axis is parallel to the rotation axis of the Earth.
We assume that a tachyon going from \emph{A} to \emph{B} or from \emph{B}
to \emph{A} is emitted as soon as a photon reaches point \emph{A}
or \emph{B}, respectively. At a given time \textit{t}, velocity \textbf{\emph{$\vec{v}$}}
($v=\beta c$, $0\leq\beta<1$) of the tachyon \textit{PF} with respect
to the laboratory frame makes the polar angle $\chi$ with the \emph{z}-axis
and the azimuthal angle $\varphi$ with the \textit{x}-axis. Due to
the Earth motion, angle $\varphi$ changes according to $\varphi\left(t\right)=\frac{2\pi}{T}t+\varphi_{0}$,
where $T$ is the sidereal day. Using the Lorentz transformations
with $\beta_{t}>1$ we find that velocity $v_{t}^{+}\left(\theta\right)$
of a tachyon going from \textit{A} to \textit{B} in the laboratory
frame is: \begin{equation}
v_{t}^{+}\left(\theta\right)=c\frac{\sqrt{\left[1+\beta\beta_{t}C\left(\theta\right)\right]^{2}+\left(\beta_{t}^{2}-1\right)\left(1-\beta^{2}\right)}}{1+\beta\beta_{t}C\left(\theta\right)}\label{eq:2}\end{equation}

where

\begin{equation}
\begin{array}{c}
C\left(\theta\right)=\left[\beta_{t}\left(1-\beta^{2}\cos^{2}\theta\right)\right]^{-1}\left[-\beta\sin^{2}\theta+\right.\\
\\\left.\cos\theta\sqrt{1-\beta^{2}}\sqrt{\left(\beta_{t}^{2}-1\right)\left(1-\beta^{2}\cos^{2}\theta\right)+1-\beta^{2}}\right].\end{array}\label{eq:3}\end{equation}
Then, velocity $v_{t}^{-}\left(\theta\right)$ of a tachyon that propagates
in the opposite direction of the \emph{x}-axis (from \emph{B} to \emph{A})
at angle $\pi-\theta$ with respect to velocity \textbf{\emph{$\vec{v}$}}
is\begin{equation}
v_{t}^{-}\left(\theta\right)=v_{t}^{+}\left(\pi-\theta\right).\label{eq:4}\end{equation}
We emphasize that velocities $v_{t}^{+}\left(\theta\right)$ and $v_{t}^{-}\left(\theta\right)$
given in eqs. \eqref{eq:2} and \eqref{eq:4} can assume negative
values as it is evident in fig. \ref{fig:DeltaSuTheta} where $\nicefrac{c}{v_{t}^{+}\left(\theta\right)}$
and $-\nicefrac{c}{v_{t}^{-}\left(\theta\right)}$ are drawn. It is
important to remark that $v_{t}^{+}\left(\theta\right)$ represents
the velocity of a tachyon going from \emph{A} to \emph{B} toward the
positive direction of the \emph{x}-axis. Since $v_{t}^{+}\left(\theta\right)=\frac{x_{B}-x_{A}}{t_{B}-t_{A}}$
and $x_{B}-x_{A}>0$, a negative value of $v_{t}^{+}\left(\theta\right)$
simply means that $t_{B}<t_{A}$, that is a tachyon emitted at \emph{A}
when the clock fixed in \emph{A} signs $t_{A}$ meets point \emph{B}
when the clock fixed in \emph{B} signs $t_{B}<t_{A}$. This is not
surprising if one takes into account that clocks at \emph{A} and \emph{B}
are synchronized using the standard Einstein synchronization procedure
and, thus, times $t_{A}$ and $t_{B}$ measured in different space
points have not an absolute meaning. For a more detailed discussion
of this point we refer to \citep{Cocciaro}.

We denote by $\Delta$ the optical path difference of entangled photons
($t_{B}=t_{A}+\nicefrac{\Delta}{c}$). Quantum correlations between
the polarizations of the entangled photons at points \emph{A} and
\emph{B} will be recovered if one of these conditions is satisfied:\textbf{\textit{}}\\
\textbf{\textit{\emph{a)}}} a tachyon emitted at \emph{A} at time
$t_{A}$ meets \emph{B} at time $t'_{B}=t_{A}+\nicefrac{d_{AB}}{v_{t}^{+}\left(\theta\right)}<t_{B}$
(before the arrival of the other entangled photon at \textit{B});\\
\textbf{\textit{\emph{b)}}} a tachyon emitted at \emph{B} at time
$t_{B}$ meets \emph{A} at time $t'_{A}=t_{B}+\nicefrac{d_{AB}}{v_{t}^{-}\left(\theta\right)}<t_{A}$
(before the arrival of the other entangled photon at A).\\
Quantum correlations will be not observed if none of these conditions
is satisfied, that is if: \begin{equation}
t_{A}+\frac{d_{AB}}{v_{t}^{+}\left(\theta\right)}>t_{A}+\frac{\Delta}{c}\qquad\vee\qquad t_{B}+\frac{d_{AB}}{v_{t}^{-}\left(\theta\right)}<t_{B}-\frac{\Delta}{c}.\label{eq:5}\end{equation}
We remark that our theoretical analysis is made under the well-founded
assumption that the polarization measurement takes place inside the
polarizing layers. From conditions \eqref{eq:5} we get:\begin{equation}
-\frac{c}{v_{t}^{-}\left(\theta\right)}<\rho<\frac{c}{v_{t}^{+}\left(\theta\right)}.\label{eq:6}\end{equation}
where $\rho=\nicefrac{\Delta}{d_{AB}}$.

Fig.\ref{fig:DeltaSuTheta} shows an example of the $\theta$-dependence
of functions $-\nicefrac{c}{v_{t}^{-}\left(\theta\right)}$ (lower
full curve) and $\nicefrac{c}{v_{t}^{+}\left(\theta\right)}$ (upper
full curve) together with function $y=\beta\cos\theta$ (broken middle
curve) for $\beta=0.5\cdot10^{-3}$ and $\beta_{t}=0.5\cdot10^{4}$.
For any value of angle $\theta$, there is always a finite interval
of values of $\rho$ (above and below $\beta\cos\theta$) that satisfies
condition \eqref{eq:6}. Using eqs. \eqref{eq:2}-\eqref{eq:5}, it
can be shown that this property remains satisfied for any $\beta<1$
and $\beta_{t}>1$. For instance, in the case shown in fig.\ref{fig:DeltaSuTheta},
no quantum correlation occurs in a finite interval of $\theta$-angles
if $-7\cdot10^{-4}<\rho<7\cdot10^{-4}$. At the first order in $\nicefrac{1}{\beta_{t}}$
inequality \eqref{eq:6} reduces to $\beta\cos\theta-\nicefrac{a\left(\beta,\theta\right)}{\beta_{t}}<\rho<\beta\cos\theta+\nicefrac{a\left(\beta,\theta\right)}{\beta_{t}}$,
where $a\left(\beta,\theta\right)=\sqrt{1-\beta^{2}}\sqrt{1-\beta^{2}\cos^{2}\theta}$.
In our experiment, angle $\theta$ between velocity \textbf{\emph{$\vec{v}$}}
of the \emph{PF} and the \emph{x}-axis (East-West direction) will
change periodically due to the Earth rotation. In particular \begin{equation}
\theta\left(t\right)=\arccos\left[-\sin\chi\sin\omega t\right]\label{eq:8}\end{equation}
where we assumed $\theta=\nicefrac{\pi}{2}$ at $t=0$ and where $\omega$
is the angular velocity of the Earth and \emph{t} is the time. Angle
$\theta$ will oscillate periodically between a minimum value $\nicefrac{\pi}{2}-\chi$
($\varphi(t)=0$ in fig.\ref{fig:ExpMeth}) and a maximum value $\nicefrac{\pi}{2}+\chi$
($\varphi(t)=\pi$ in fig.\ref{fig:ExpMeth}). Looking at fig.\ref{fig:DeltaSuTheta}
we see that, for $\Delta=0$, there exists always a finite interval
of angles where no quantum correlation must occur. Then, it is convenient
to set $\Delta=0$ in the experiment. However, the presence of an
experimental uncertainty $\delta\Delta$ sets severe limitations to
the possibility of observing a lack of quantum correlation.

In our experiment, distance $d_{AB}$ between the polarizer layers
at \emph{A} and \emph{B} is $d_{AB}=1.75\, m$ and the main experimental
uncertainties are:\\
\textbf{1)} the uncertainty on the positioning of polarizers at
the same distance from the source of entangled photons ($\delta\Delta\approx40\,\mu m$);\\
\textbf{2)} the intrinsic uncertainty related to the coherence
length $h_{c}=\nicefrac{\lambda^{2}}{\Delta\lambda}$ of the entangled
photons. The entangled photons pass through a $40\, nm$ width optical
filter centered at $820\, nm$ leading to $h_{c}\approx16\,\mu m$;\\
\textbf{3)} the spatial width of the active polarizing layers
that is $d_{c}\approx220\,\mu m$.

The resulting uncertainty is $\delta\Delta\approx280\,\mu m$, then,
in our experiment

\begin{equation}
\rho=0\pm\bar{\rho}=\pm1.6\cdot10^{-4},\label{eq:9}\end{equation}
where \begin{equation}
\bar{\rho}=\frac{\delta\Delta}{d_{AB}}.\label{eq:9-1}\end{equation}
A total lack of quantum correlation will be only observed if angle
$\theta$ is within the interval $\left[\theta_{1},\theta_{2}\right]$
of amplitude $\Delta\theta$ shown in fig.\ref{fig:DeltaSuTheta}.
Since $\theta=\theta\left(t\right)$, the condition $\theta_{1}<\theta<\theta_{2}$
establishes a corresponding time interval $\Delta t$ {[}$\theta\left(\nicefrac{\Delta t}{2}\right)-\theta\left(\nicefrac{-\Delta t}{2}\right)=\theta_{2}-\theta_{1}${]}.
Then,

\begin{equation}
\cos\theta_{1,2}=\cos\left[\theta\left(\mp\frac{\Delta t}{2}\right)\right]=-\sin\chi\sin\left(\mp\frac{\omega\Delta t}{2}\right),\label{eq:10}\end{equation}
where symbols - and + refer to $\theta_{1}$ and $\theta_{2}$, respectively.
The maximum value of interval $\Delta\theta$ for which a total lack
of correlation can be observed is fixed by $\bar{\rho}$ and will
be obtained using the condition \begin{equation}
2\bar{\rho}<\frac{c}{v_{t}^{+}\left(\theta_{2}\right)}+\frac{c}{v_{t}^{-}\left(\theta_{1}\right)}\label{eq:11-1}\end{equation}
that corresponds to impose that the rectangle of sides $\Delta\theta$
and $2\bar{\rho}$ is fully inside the region limited by the curves
$\nicefrac{c}{v_{t}^{+}\left(\theta\right)}$ and $\nicefrac{-c}{v_{t}^{-}\left(\theta\right)}$
(see the gray rectangle in fig.\ref{fig:DeltaSuTheta}). Substituting
the expressions of $v_{t}^{+}\left(\theta\right)$ and $v_{t}^{-}\left(\theta\right)$
given in eqs. \eqref{eq:2}-\eqref{eq:4} into \eqref{eq:11-1} with
$\cos\theta_{1}$ and $\cos\theta_{2}$ given by \eqref{eq:10}, after
some tedious but straightforward calculations we finally obtain:\begin{equation}
\beta_{t}<\beta_{t,min}=\sqrt{1+\frac{\left(1-\beta^{2}\right)\left[1-\bar{\rho}^{2}\right]}{\left[\bar{\rho}+\beta\sin\chi\sin\frac{\omega\Delta t}{2}\right]^{2}}}.\label{eq:11-2}\end{equation}
Equation \eqref{eq:11-2} was already obtained by Salart et al.\citep{Salart}
and establishes the basic dependence of the maximum detectable tachyon
velocity $\beta_{t,min}$ on the experimental parameters $\bar{\rho}=\nicefrac{\delta\Delta}{d_{AB}}$
and $\Delta t$. For a fixed value of $\beta$, $\beta_{t,min}$ decreases
increasing both $\bar{\rho}$ and $\Delta t$ ($\Delta t<\nicefrac{\pi}{\omega}$).

In conclusion, in our experiment, whatever is velocity \textbf{\emph{$\vec{v}$}}
(modulus $\beta c$ and direction $\chi$) of the \textit{PF}, deviations
from the predictions of the Quantum Mechanics should be always observed
provided the tachyon velocity in the \textit{PF} satisfies condition
\eqref{eq:11-2}.

\section{\label{sec:--Experimental-apparatus}- Experimental apparatus and
procedures.}

\begin{figure}[tbh]
\begin{centering}
\includegraphics[bb=0bp 0bp 539bp 500bp,scale=0.25]{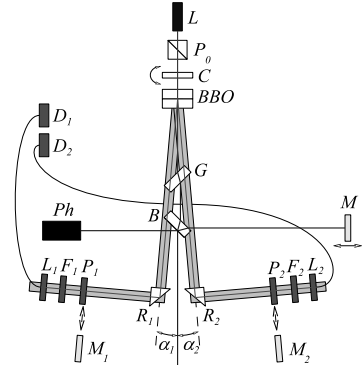}
\par\end{centering}

\caption{\label{fig:ExpApp}schematic view of the experimental apparatus}

\end{figure}

The experimental apparatus is schematically drawn in fig.\ref{fig:ExpApp}.
A blue diode laser beam (\textit{L}, $\lambda=407\, nm$) is polarized
at 45\textdegree{} with respect to the vertical axis by polarizer
$P_{0}$, passes through a tilting plate compensator (\textit{C})
with vertical extraordinary axis and impinges at normal incidence
on two thin ($0.5\, mm$) adjacent nonlinear optical crystals (\textit{BBO})
cut for type-I phase matching. The two crystals are aligned so that
their optic axes lie in planes perpendicular to each other with the
first plane that is horizontal. The coherence length of the blue laser
beam was increased to $1\, mm$ using a reflection diffraction grating
that is not shown in fig.\ref{fig:ExpApp}. Since the pump polarization
is set at 45\textdegree{}, it induces down conversion at $\lambda=814\, nm$
in each crystal \citep{Kwiat}. The output photons are created in
the maximally entangled state $\left(|H,H>+e^{i\phi}|V,V>\right)/\sqrt{2}$,
where phase $\phi$ can be adjusted via tilting of the optical compensator
\textit{C}. The remaining components ($R_{i},P_{i},F_{i},L_{i}$ and
$D_{i}$) are mounted on two \textit{L}-shaped guides that can rotate
in the horizontal plane of the optical table around a vertical axis
passing for the center of the \textit{BBO} plates so that angles $\alpha_{1}$
and $\alpha_{2}$ can be changed continuously with an accuracy of
0.02\textdegree{}. $R_{1}$ and $R_{2}$ are two right angle prisms,
$P_{1}$ and $P_{2}$ are thin polarizing films (LPNIR, Thorlabs),
$F_{1}$ and $F_{2}$ are interference filters ($\lambda=820\, nm\pm20\, nm$)
and $D_{1}$ and $D_{2}$ are single photons counters (Perkin Elmer
SPCM-AQ4C). $L_{1}$ and $L_{2}$ are optical lenses that focus the
entangled photons on two multimode fibers connected to detectors $D_{1}$
and $D_{2}$. The detectors outputs are sent to electronic counters
and to a coincidence circuit and, after, to a \textit{PC}. A labview
program controls any experimental feature. The \textit{L}-shaped guides
are rotated to maximize the count rates and the coincidences of the
entangled photons. This condition is satisfied for $\alpha_{1}=\alpha_{2}=2\text{\textdegree}$.
In this condition, the centers of polarizers $P_{1}$ and $P_{2}$
are aligned along a \textit{x}-axis in the East-West direction within
0.2\textdegree{}. To minimize any possible thermal displacement, the
room temperature is held fixed at $27\text{\textdegree}\pm0.3\text{\textdegree}C$
during the whole measurements time.

In fig.\ref{fig:ExpApp} are also shown a glass plate \textit{G},
a beam splitter \textit{B}, a reference mirror \textit{M} and a photodiode
\textit{Ph}. These components are introduced only during the preliminary
measurements to equalize the optical paths 1 and 2 of entangled photons
from the \textit{BBO} plates to the first surfaces of polarizers $P_{1}$
and $P_{2}$ using an interferometric method. Glass plate \textit{G}
is introduced to compensate the lateral shift of the blue laser beam
due to beam splitter \textit{B}. \textit{G} and \textit{B} are mounted
in such a way that the rotation of the \textit{L}-shaped guides is
not disturbed. The first step consists on replacing polarizer $P_{1}$
by a mirror $M_{1}$, then the left \textit{L}-shaped guide is rotated
up to $\alpha_{1}=0$. In this condition, paths $B-R_{1}-M_{1}$ and
\textit{$B-M$}, constitute the two arms of a Michelson-like interferometer
for the blue laser beam. Then, interference fringes occur in front
of photodiode \textit{Ph}. Due to the finite coherence length of our
blue laser beam (about $1\, mm$), the maximum contrast of interference
fringes is obtained for equal lengths of the two arms of the interferometer.
To measure the fringe contrast, we put a small loudspeaker oscillating
at a frequency of $100\, Hz$ in the contact with the reference arm
of the interferometer containing mirror \emph{M}. This procedure induces
small changes ($\lesssim1\,\mu m$) of the length of the reference
arm and, thus, a corresponding oscillation of the pattern of interference
fringes in front of photodiode \textit{Ph}. Then, the amplitude of
the oscillating signal at the output of photodiode \textit{Ph} is
proportional to the fringe contrast. The reference mirror \textit{M}
is mounted on a motorized linear stage and the mirror position leading
to the maximum amplitude of the oscillating signal is found with an
estimated precision better than $5\,\mu m$. The same kind of measurement
is, then, repeated with the right \textit{L}-shaped guide after polarizer
$P_{2}$ has been replaced by mirror $M_{2}$. In this latter case,
the motorized linear stage of the reference mirror \textit{M} is held
at the rest while $M_{2}$ is translated up to recover the maximum
contrast of interference fringes. In these conditions the optical
paths of entangled photons 1 and 2 (up to mirrors $M_{1}$ and $M_{2}$)
are the same within $10\,\mu m$. Due to thermal effects, small changes
of the lengths of the two arms could occur. To take under control
this possible effect, we have repeated the measurement during a week
at different times measuring the variations of differences of lengths
of the interferometer arms. These variations are usually smaller than
$5\,\mu m$ but, in a few cases, changes up to $25\,\mu m$ have been
observed. The final procedure consists on replacing mirrors $M_{1}$
and $M_{2}$ with polarizers $P_{1}$ and $P_{2}$ in such a way that
the first surface of the polarizing films (firstly encountered by
the entangled photons) lies in the same position of the mirror surfaces.
This is obtained using a profilometer that ensures a final accuracy
better than $5\,\mu m$. Therefore, tacking into account for possible
thermal drifts, the resulting difference of optical paths between
the entangled photons in the two arms is estimated to be less than
$40\,\mu m$.

Once the two optical paths have been equalized, the components \textit{G},
\textit{B}, \textit{M} and \textit{Ph} are removed and plate \textit{G}
is replaced by a black light absorber, then the \textit{L}-shaped
guides are positioned at angles $\alpha_{1}=\alpha_{2}=2\text{\textdegree}$
where the maximum of coincidences is observed. We denotes here by
$\gamma_{1}$ and $\gamma_{2}$ the angles between the polarization
axes of polarizers $P_{1}$ and $P_{2}$ and the vertical axis. According
to \textit{QM}, the probability of coincident detection of photons
passing through the polarizers is

\begin{equation}
P_{12}(\gamma_{1},\gamma_{2})=\frac{\left|\cos\gamma_{1}\cos\gamma_{2}+e^{i\phi}\sin\gamma_{1}\sin\gamma_{2}\right|^{2}}{2}\label{eq:12}\end{equation}
whilst the probability of detection of a photon passing through a
single polarizer is

\begin{equation}
P_{1}(\gamma_{1})=P_{2}(\gamma_{2})=\frac{1}{2}.\label{eq:13}\end{equation}
In our experiment we set $\gamma_{1}=\gamma_{2}=\nicefrac{\pi}{4}$
and, thus, from eq.\eqref{eq:12} we infer that the number of coincidences
depends on phase $\phi$ according to

\begin{equation}
n_{coinc}(\phi)=n_{max}\cos^{2}\frac{\phi}{2}.\label{eq:14}\end{equation}
The result in eq.\eqref{eq:14} is predicted by \textit{QM} and by
the Eberhard model if there has been sufficient time for communication
between photons 1 and 2. On the contrary, if communication is not
possible (see section \ref{sec:Conditions-for-the}), the Eberhard
model predicts that the measurements at points \textit{A} and \textit{B}
are fully uncorrelated (eqs. (4.68) and (4.81) in \citep{Eberhard})
and the probability of coincident detection of photons becomes $P_{12}(\gamma_{1},\gamma_{2})=P_{1}(\gamma_{1})P_{2}(\gamma_{2})=\nicefrac{1}{4}$
that leads to the constant coincidences rate

\begin{equation}
n_{coinc}^{*}=\frac{n_{max}}{2}.\label{eq:15}\end{equation}
If $\phi=\pi$, the \textit{QM} coincidence rate predicted by eq.\eqref{eq:14}
becomes $n_{coinc}=0$. Then, according to the Eberhard model, a sharp
variation from the \textit{QM} value $n_{coinc}=0$ to the uncorrelated
one $n_{coinc}^{*}=\nicefrac{n_{max}}{2}$ should be observed in a
suitable time interval of a sidereal day if $\phi=\pi$ and $\beta_{t}<\beta_{min}$
given in eq.\eqref{eq:11-2}\textit{.}

\section{- \label{sec:--Experimental-results}Experimental results and conclusions.}

\begin{figure}[tbh]
\begin{centering}
\includegraphics[scale=0.3]{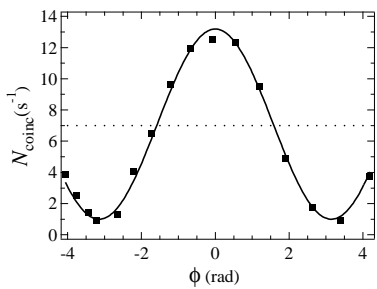}
\par\end{centering}

\caption{\label{fig:Phase}Coincidence rate (counts/s) versus the average phase
$\phi$. The acquisition time is $\Delta t=100\, s$. The full line
is the best fit with function $N_{coinc}=a+b\cos^{2}\left(\frac{\phi}{2}\right)$
with $a=1.14\, s^{-1}$ and $b=12.05\, s^{-1}$. The dotted horizontal
line represents the theoretical prediction of the Eberhard model if
quantum communication are not allowed (eq. \eqref{eq:15}).}

\end{figure}

Fig.\ref{fig:Phase} shows the experimental values of the coincidence
rate ($N_{coinc}=\nicefrac{n_{coinc}}{\Delta t}$) versus phase $\phi$
that are obtained using a $100\, s$ acquisition time and setting
the polarizers angles at $\gamma_{1}=\gamma_{2}=\nicefrac{\pi}{4}$.
The full line represents the best fit with function $N_{coinc}=a+b\cos^{2}(\phi/2)$
that differs from the theoretical prediction in eq.\eqref{eq:14}
due to the presence of constant \textit{a}. The dotted horizontal
line represents the theoretical prediction for no quantum communication.
The non-vanishing value of coefficient \textit{a} is due in part to
the dark spurious coincidences ($N_{dark}=0.3\, s^{-1}$) and in part
($0.84\, s^{-1}$) to the finite acceptance angle of entangled photons
that is estimated to be $\Delta\alpha\approx5\cdot10^{-3}\, rad$
in our experiment. In fact, due to the birefringence of \emph{BBO},
the phases differences between entangled photons that are emitted
at slightly different angles have not the same values leading to the
reduced contrast observed in fig.\ref{fig:Phase} \citep{Kwiat2}.

In order to detect a possible lack of quantum correlation occurring
at a given time interval during a sidereal day, we set $\gamma_{1}=\gamma_{2}=\nicefrac{\pi}{4}$
and $\phi=\pi$. Indeed, according to eqs. \eqref{eq:14} and \eqref{eq:15},
the maximum departure between quantum correlated and uncorrelated
results (full and dotted lines in fig.\ref{fig:Phase}) is expected
if phase $\phi$ is a multiple of $\pi$. We choose an odd multiple
of $\pi$ since it corresponds to a minimum of the coincidence number
$n_{coinc}$ and, thus, to a minimum value of the statistical noise
$\sqrt{n_{coinc}}$. According to eq.\eqref{eq:11-2}, the higher
value of the lower limit $\beta_{min}$ would be obtained for $\Delta t\rightarrow0$.
However, a reduction of the acquisition time leads to a decrease of
the signal to noise ratio. In our experiment, sufficient signal to
noise ratio is obtained setting $\Delta t=4\, s$. With this choice,
$\beta\sin\chi\sin\frac{\omega\Delta t}{2}<\bar{\rho}$ for any value
of $\beta$ and $\chi$ and, thus, the finite acquisition time affects
appreciably $\beta_{min}$ only for relativistic values of $\beta$.
Using this acquisition time, we measure the number of coincidences
at different times during a sidereal day.%
\begin{figure}[tbh]
\begin{centering}
\includegraphics[scale=0.3]{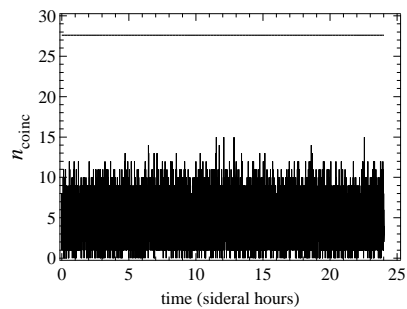}
\par\end{centering}

\caption{\label{fig:OneDay}Number of coincidences versus time expressed in
sidereal hours. The acquisition time is $\Delta t=4\, s$. The full
line is the number of coincidences predicted for no quantum correlation
between entangled photons.}

\end{figure}
\begin{figure}[tbh]
\begin{centering}
\includegraphics[scale=0.3]{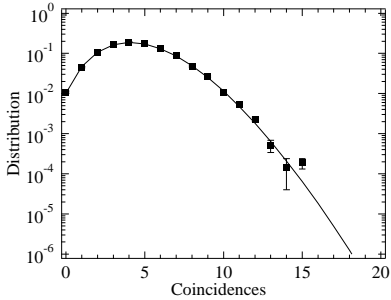}
\par\end{centering}

\caption{\label{fig:Distr}Relative frequency distribution of coincidences
in a simple sidereal day for a $4\, s$ acquisition time. The experimental
data are those of fig.\ref{fig:OneDay}. The full line represents
the theoretical Poisson distribution.}

\end{figure}
The results of this measurement are shown in fig.\ref{fig:OneDay}
where the horizontal line represents the value of coincidences expected
for totally uncorrelated photons {[}$n_{coinc}=\left(a+\nicefrac{b}{2}\right)\Delta t${]}.
Although the experimental noise is somewhat high due to the small
acquisition time, the coincidences rate remains always well below
the horizontal line. Furthermore, the fluctuations of the coincidences
around the average value $n_{av}=4.58$ that are clearly visible in
fig.\ref{fig:OneDay} are fully consistent with the usual Poisson
noise. This is evident in fig.\ref{fig:Distr} where the relative
frequency distribution of coincidences during a sidereal day is shown
together with the Poisson distribution (full line) that corresponds
to $n_{av}=4.58$. We emphasize that no free parameter is used to
draw the full line in fig.\ref{fig:Distr}.

\begin{figure}[tbh]
\begin{centering}
\includegraphics[scale=0.3]{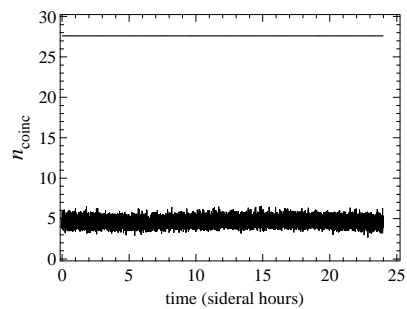}
\par\end{centering}

\caption{\label{fig:21Days}Number of coincidences with the acquisition time
$\Delta t=4\, s$ averaged over 21 sidereal days versus time.}

\end{figure}

According to Eberhard, the model in \citep{Eberhard} is only one
of the possible realistic local models based on signals propagating
at a superluminal velocity that can be proposed to justify the observed
quantum correlations in EPR experiments. In particular, in such a
model, the correlation between entangled photons is entirely due to
the quantum communication and, thus, no correlation exists if they
have not sufficient time to communicate. However, more complex models
that are a combination of hidden variable theories and quantum communications
could be proposed \citep{Bohm}. According to these models, the quantum
correlation between entangled photons could be due in part to some
correlation that is already present at the beginning when entangled
particles are created (hidden variables) and in part to superluminal
communications between them. In such a case, some residual correlation
of entangled photons would be present also if they have not sufficient
time to communicate and, thus, the number of coincidences that should
be expected when there is no communication could be appreciably different
from the value $n_{max}/2$ that characterizes totally uncorrelated
events {[}$P_{12}(\gamma_{1},\gamma_{2})\neq P_{1}(\gamma_{1})P_{2}(\gamma_{2})=\nicefrac{1}{4}${]}.
In such a case, one or more of the coincidences variations shown in
fig.\ref{fig:OneDay} could be not due to the statistical noise but
to the occurrence of some incomplete quantum correlation at a given
time interval. Although this interpretation seems to be reasonably
excluded by the satisfactory agreement between the experimental frequency
distribution of the coincidences and the Poisson distribution (fig.\ref{fig:Distr}),
we decided to improve our analysis in order to evidence possible small
systematic deviations. In order to reach this goal, we exploit an
important feature of the Eberhard model: the lack of quantum correlation
is expected to be a periodic phenomenon with a period of a sidereal
day (in the ideal case $\rho=0$, the period becomes a half of a sidereal
day). Then, a small systematic signal occurring at a given time of
a sidereal day could be enhanced with respect to the statistical noise
by summing the counts occurring at the same instant of $N$ different
sidereal days. This procedure reduces the statistical noise by a factor
$\sqrt{1/N}$ but does not affect a possible periodic signal. For
this reason we have repeated measurements over 21 sidereal days. The
counts corresponding to the same time intervals of different sidereal
days have been mediated. The results of this procedure are shown in
fig.\ref{fig:21Days}. Now, the statistical noise is appreciably reduced
and the difference between the experimental counts and those expected
for no correlated events (full horizontal line) is much more evident.
Furthermore, no peak comes out from the statistical noise. In particular,
the maximum departure of $n_{coinc}$ from the average value $n_{av}=4.62$
is $\Delta n=1.95$ (the standard deviation in fig.\ref{fig:21Days}
is $\sigma=0.48$). Such a variation corresponds to about 1/12 of
the expected variation $\Delta n=23$ for totally uncorrelated results.%

Substituting our experimental parameters $\bar{\rho}=1.6\cdot10^{-4}$
and $\Delta t=4\, s$ in eq.\eqref{eq:11-2} we can calculate the
lower bound $\beta_{t,min}$ versus $\beta$ and $\chi$. %
\begin{figure}[tbh]
\begin{centering}
\includegraphics[scale=0.3]{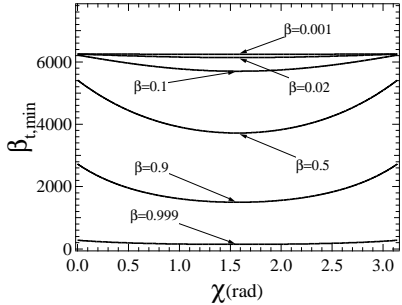}
\par\end{centering}

\caption{\label{fig:BetaMaxSuChi} $\beta_{t,min}$ versus $\chi$ for some
values of $\beta$ obtained using eq.\eqref{eq:11-2} with the experimental
parameters $\bar{\rho}=1.6\cdot10^{-4}$ and $\Delta t=4\, s$.}

\end{figure}
\begin{figure}[tbh]
\begin{centering}
\includegraphics[scale=0.3]{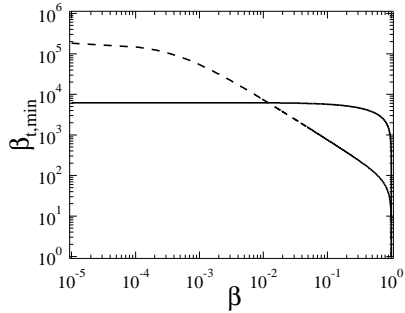}
\par\end{centering}

\caption{\label{fig:BetaMaxSuBeta} $\beta_{t,min}$ versus $\beta$ for $\chi=\nicefrac{\pi}{2}$.
The full line is obtained for $\bar{\rho}=1.6\cdot10^{-4}$ and $\Delta t=4\, s$
using eq.\eqref{eq:11-2} whilst the broken line corresponds to the
parameters of the experiment in ref. \citep{Salart} ($\bar{\rho}=5.4\cdot10^{-6}$
and $\Delta t=360\, s$).}

\end{figure}
Fig.\ref{fig:BetaMaxSuChi} shows the dependence of $\beta_{t,min}$
on $\chi$ for some values of $\beta$. Note that $\beta_{t,min}$
is virtually independent of $\chi$ for $\beta\lesssim10^{-2}$ and
that, for a fixed $\beta$, the minimum value of $\beta_{t,min}$
is reached for $\chi=\nicefrac{\pi}{2}$. The relatively smooth dependence
of $\beta_{t,min}$ on $\chi$ is the direct consequence of two main
features of our experiment: a) the choice of a East-West orientation
of the axis connecting the polarizers; b) the choice of a relatively
small $\Delta t$ {[}$\sin\left(\nicefrac{\Delta t}{2}\right)\lesssim\bar{\rho}$
in eq.\eqref{eq:11-2}{]}. In particular, if the detection axis makes
a finite angle $\alpha$ with the East-West direction, a sharp decrease
of $\beta_{t,min}$ occurs for $0\leq\chi\lesssim\alpha$ and for
$\pi-\alpha\lesssim\chi\leq\pi$ (see fig. 5a of \citep{Salart}).
Fig.\ref{fig:BetaMaxSuBeta} shows the dependence of $\beta_{t,min}$
on $\beta$ for $\chi=\nicefrac{\pi}{2}$. The full line corresponds
to $\beta_{t,min}$ as obtained using the parameters $\bar{\rho}=1.6\cdot10^{-4}$
and $\Delta t=4\, s$ of our experiment whilst the broken line is
obtained using the parameters $\bar{\rho}=5.4\cdot10^{-6}$ and $\Delta t=360\, s$
of ref. \citep{Salart}. Note that our experimental values of $\beta_{t,min}$
are virtually independent of $\beta$ up to $\beta\approx0.1$ ($\beta_{t,min}\gtrsim0.6\cdot10^{4}$
for $\beta<0.1$). This is another direct consequence of relatively
small value of $\Delta t$. The experimental results of Salart et
al. \citep{Salart} (broken line in fig. \ref{fig:BetaMaxSuBeta})
show a much greater dependence on $\beta$ since $\sin\left(\nicefrac{\Delta t}{2}\right)\gg\bar{\rho}$
in their experiment. Note that the values of $\beta_{t,min}$ of the
broken curve are more than one order of magnitude higher than those
of the full curve if $\beta\lesssim10^{-3}$ and become appreciably
smaller for $\beta\gtrsim10^{-2}$. We remind here that the velocity
of the cosmic microwave background radiation reference frame is $v\approx1.2\cdot10^{-3}\, c$
\citep{Scarani}. Finally, we emphasize that, for any value of $\Delta t$
and $\bar{\rho}$, $\beta_{t,min}\rightarrow1$ for $\beta\rightarrow1$.

In conclusion, in this Letter we have experimentally investigated
the possibility that the quantum correlation that characterizes EPR
experiments can be due to some superluminal signal propagating in
a preferred frame. Due to the choice of a West-East alignment of the
polarizers and to the choice of a sufficiently small acquisition time,
our experimental results are poorly dependent on both the modulus
and the orientation of the velocity of the preferred frame. No evidence
for the presence of superluminal signals has been observed in our
experiment. Our experimental results provide a lower bound for the
velocity of possible superluminal signals as shown in the full curve
of fig.\ref{fig:BetaMaxSuBeta}.

\section*{Acknowledgments}

The authors greatly acknowledge Marco Bianucci for valuable technical
support, Giuseppe Cocciaro for the support given at the beginning
of the experiment and Nicolas Gisin for critical comments.

\end{document}